\documentclass[
               twocolumn,
               noshowpacs,          
               nopreprintnumbers,     
               aps,                 
               prd,                 
               a4paper,             
               superscriptaddress,  
               nofootinbib,         
               tightenlines,        
               floats,floatfix      
               ]{revtex4}

\usepackage{amsmath, amsfonts, amsthm, amssymb, graphicx}
\usepackage{epstopdf}
 \usepackage{slashed}
 \usepackage{graphicx,epsfig}
\usepackage{amsmath}
\usepackage {amssymb}
\usepackage[utf8]{inputenc}
\usepackage{hyperref}
\usepackage[english]{babel}

\newcommand{\be}{\begin{eqnarray}}
\newcommand{\ee}{\end{eqnarray}}
\newcommand{\bea}{\begin{eqnarray}}
\newcommand{\eea}{\end{eqnarray}}

\begin{document}

\title{Sub-annular structure in black hole image from gravitational refraction}

\author{Gaston Giribet}
\affiliation{Department of Physics, New York University, 726 Broadway, New York, NY10003, USA.}

\author{Emilio Rub\'{\i}n de Celis}
\affiliation{Departamento de F\'isica, Universidad de Buenos Aires FCEN-UBA and IFIBA-CONICET, Ciudad Universitaria, Pabell\'on 1, 1428, Buenos Aires, Argentina.}

\author{Pedro Schmied}
\affiliation{Departamento de F\'isica, Universidad de Buenos Aires FCEN-UBA and IFIBA-CONICET, Ciudad Universitaria, Pabell\'on 1, 1428, Buenos Aires, Argentina.}



\begin{abstract}
The images of supermassive black holes captured by the Event Horizon Telescope (EHT) collaboration have allowed us to have access to the physical processes that occur in the vicinity of the event horizons of these objects. This has enabled us to learn more about the state of rotation of black holes, about the formation of relativistic jets in their vicinity, about the magnetic field in the regions close to them, and even about the existence of the photon ring. Furthermore, black hole imaging gives rise to a new way of testing general relativity in the strong field regime. This has initiated a line of research aimed at probing different physical scenarios. While many scenarios have been proposed in the literature that yield distortion effects that would be a priori detectable at the resolution achieved by future EHT observations, the vast majority of those scenarios involve strange objects or exotic matter content. Here, we consider a less heterodox scenario which, involving non-exotic matter, in the sense that it satisfies all energy conditions and is dynamically stable, also leads to a deformation of the black hole shadow. We consider a specific concentration of non-emitting, relativistic matter of zero optical depth forming a bubble around the black hole. Due to gravitational refraction, such a self-interacting --dark-- matter concentration may produce sub-annular images, i.e. subleading images inside the photon ring. We calculate the ray tracing in the space-time geometry produced by such a matter configuration and obtain the corresponding black hole images. While for concreteness we restrict our analysis to a specific matter distribution, modeling the bubble as a thin-shell, effects qualitatively similar to those described here are expected to occur for more general density profiles. 

\end{abstract}

\maketitle

\section{Introduction} 

\begin{figure}
    \centering
\includegraphics[width=0.47\textwidth]{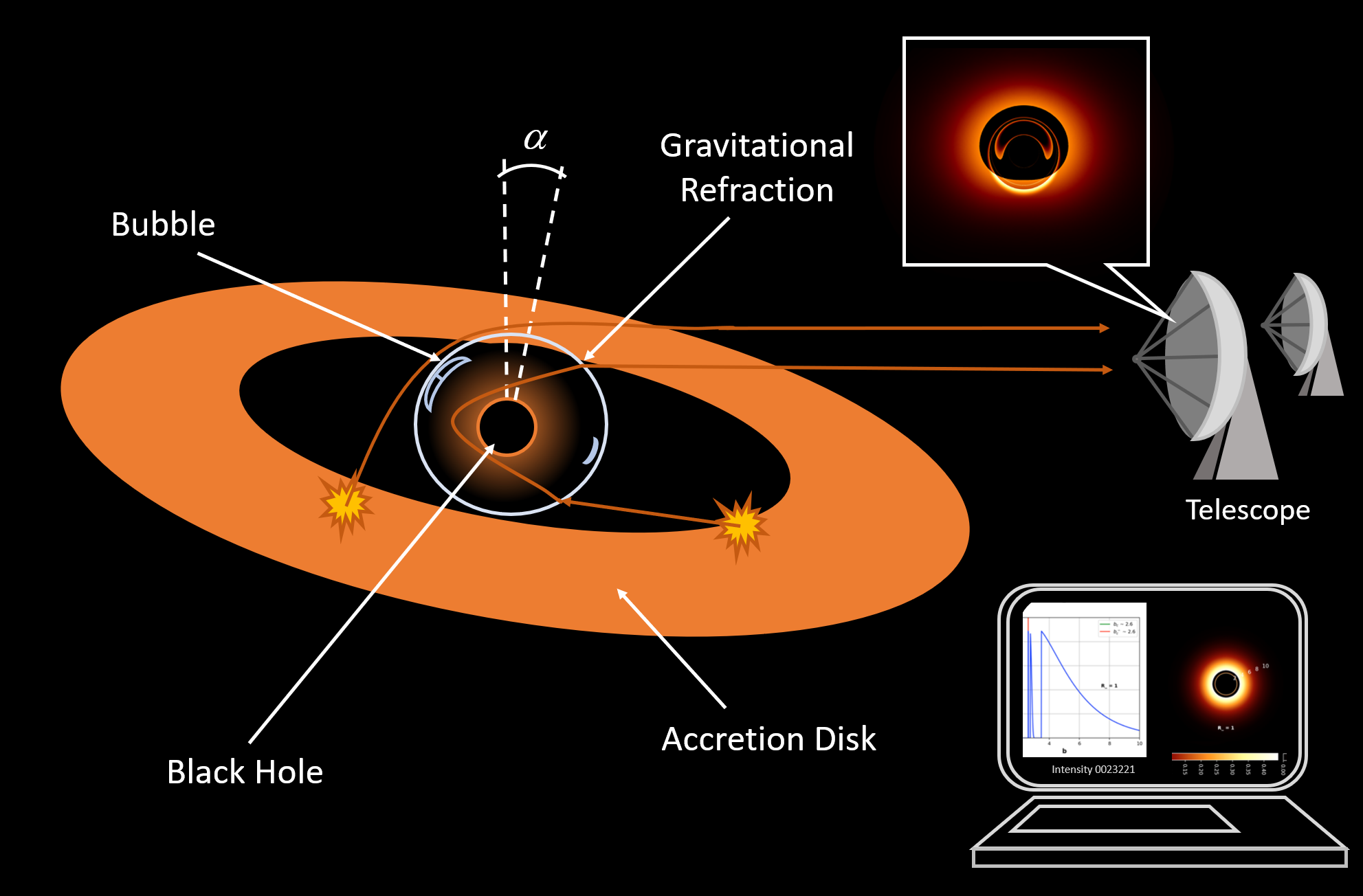}
    \caption{Scheme of the physical configuration, consisting of a black hole surrounded by a bubble of dark matter and an optically thin accretion disk. Light rays experience gravitational refraction, what results in the formation of sub-annular images, i.e. images inside the photon ring.}
    \label{fig:esquema.png}
\end{figure}

This is an exciting time for the study of black holes. The recent observation of the silhouettes of supermassive black holes, by the Event Horizon Telescope collaboration, has marked the beginning of a new era in the research of these objects. 
\begin{figure}
    \centering
\includegraphics[width=0.30\textwidth]{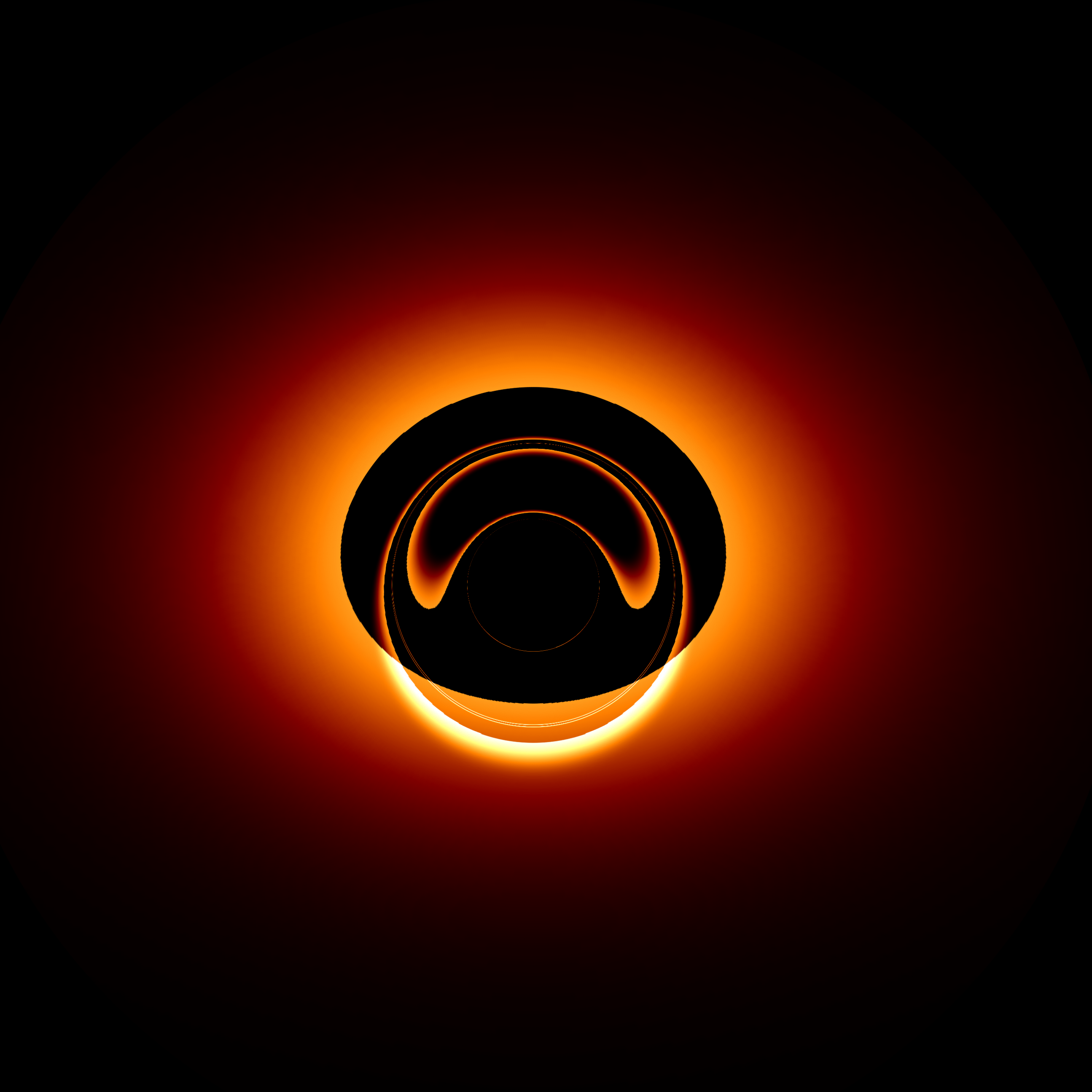}
    \caption{Image of the system with a black hole of radius ${R_- = 0.3}$ and the thin-shell located in ${R = 1.5}$, when considering an accretion disk with a $45^o$ inclination respect to the observer.}
\label{fig:inclinado_ejemplo_inicial.png}
\end{figure}
The observation of the image of the supermassive black hole in the core of galaxy Messier 87 \cite{EHT_I} and in the center of our own galaxy \cite{EHT_Sag_I} makes it clear that, soon, with the advent of more radio telescopes, new analysis techniques and expansion of the radio frequency range, we will have access to regions that are actually near the event horizon of these objects. In the last four years, images have shown us fine details of structure of black hole jets \cite{Jet_Centauro_A}, of the polarization produced by the magnetic fields in which they are embedded \cite{EHT_VII, EHT:2023thr}, and of the abstruse accretion mechanisms that take place in their vicinity \cite{Lu:2023bbn}. Coming observations will surely teach us more about the dynamics of accretion processes, as well as about the state of rotation of black holes, and perhaps about the very space-time geometry near them. The possibility of scrutinizing the shape of the shadow of black holes with precision opens the door to new ways of testing general relativity in the strong field regime \cite{Amarilla:2010zq, Psaltis_Effects_Spacetime_Geometry, Psaltis_test_GR_Plasma, EHT_charges, Psaltis_post_newtonian_shadow_test, Glampedakis:2021oie}. This led some to speculate that studying the shadows of black holes with enough detail could be used to discern whether the objects found in galactic centers are actually black holes or, in contrast, some other type of objects predicted by speculative theories. This initiated an ample line of research aimed at predicting the shadow produced by all sort of exotic objects, including wormholes \cite{Nedkova:2013msa}, gravastars \cite{Gravitational_lensing_gravastars}, topological stars \cite{Topological_stars_lensing}, higher-dimensional black holes \cite{Amarilla:2011fx, Amarilla:2013sj}, naked singularities \cite{Naked_Singularity_1, Naked_Singularity_2, naked_sing_lensing}, among others. Here, we will consider a less heterodox scenario, which, while still producing a distortion of the black hole image, consists of a particular matter concentration that is not exotic, in the sense that it respects all the energy conditions and turns out to be dynamically stable. This comprises a static black hole surrounded by a bubble made of non-emitting matter of zero optical depth. While the phenomenon we will describe occurs in the case of more general distributions profiles of such --dark-- matter, as a working example we will focus on the particular case in which the bubble is modeled by a thin-shell of transparent matter encapsulating the black hole. This example will suffice to show that such a configuration can be stable for a physically sensible set of parameters and, due to gravitational refraction, produce an observable distortion in the rings pattern of the image \cite{Broderick:2022tfu}, i.e. the distribution pattern of gravitationally lensed secondary images produced by the photon spheres; see Figure \ref{fig:inclinado_ejemplo_inicial.png}. This is sufficient to produce a sub-annular structure, inside the photon ring, without the requirement of exotic matter or strange spacetime contortions. The physical scenario is schematically depicted in Figure \ref{fig:esquema.png}. Figure \ref{fig:inclinado_ejemplo_inicial.png} shows the predicted shadow for such a configuration, with the sub-annular structure being due to the gravitational refraction produced by the bubble.

\section{Physical scenario} \label{construccion_sistema}
The configuration we will consider consists of a static black hole surrounded by a thin shell bubble made of non-emitting, transparent matter. The thin-shell is located at a fixed radial distance $R$ from the black hole, somewhere between the photon sphere and the innermost stable circular orbit (ISCO). We demand the configuration to be dynamically stable and obey both the strong and the dominant energy conditions. That is to say, while the induced stress-tensor on the bubble needs to exhibit a large skin tension in order to sustain the configuration and prevent gravitational collapse, the induced stress tensor on the bubble will still be demanded to have positive energy density and causal --sub-luminal-- acoustic excitations. Remarkably, general relativity allows for configurations of that sort which, in addition, turn out to be stable under radial perturbations.

The constant-time sections of the world-volume of the thin-shell are codimension-one spacelike surfaces embedded in a (3+1)-dimensional spacetime composed by two different patches, the interior patch $\mathcal{M}^-$ and the exterior patch $\mathcal{M}^+$. In both patches the spacetime metric is that of Schwarzschild, with different mass parameters $m_{\pm}$ each; while the metric in $\mathcal{M}^-$ only accounts for the mass of the black hole, the mass parameter of the exterior metric, $m_+$, also accounts for the energy density of the bubble,  $\sigma $. We will denote $R_\pm = 2m_{\pm }G/c^2$ the horizon radius of the metric on $\mathcal{M}^{\pm}$, and $R$ indicates the radial location of the shell (with $G$ and $c$ being the gravitational constant and the speed of light, respectively). In other words, $R_-$ is the radius of the event horizon of the black hole; we will have $R>R_+$ and consider units such that $R_+=1$.

Solving Einstein equations through the thin-shell amounts to imposing the Israel junction condition, which takes the form
\begin{align}\label{eq=Lanczos}
    t_{i j}\,=\,\frac{c^4}{8 \pi G}(\,[K]\,h_{i j}\,-\,[K_{i j}]\,)\, ,
\end{align}
where $t_{ij}$ and ${h_{ij}}$ are the induced stress-tensor and the induced metric on the shell, respectively; ${[K_{ij}]\,=\,K_{ij}^+\,-\,K_{ij}^-}$ is the discontinuity of the extrinsic curvature, with ${[K]=h^{ij}[K_{ij}]}$. Following a cut and paste procedure, we take the two incomplete copies of Schwarzschild spacetime, $\mathcal{M}^-$ and $\mathcal{M}^+$,  and connect them. The entire spacetime ${\mathcal{M}\,=\,\mathcal{M}^+\,\cup\,\mathcal{M}^-}$ is constructed by identifying the respective boundaries $\partial\mathcal{M}^-=\partial\mathcal{M}^+$ across the spherically symmetric surface located at radius $R$ from the center of $\mathcal{M}^-$. The resulting spacetime turns out to be geodesically complete everywhere outside and on the black hole event horizon.

From the junction condition (\ref{eq=Lanczos}) we can derive an expression for the energy density $\sigma$ and the surface pressure $p$ on the bubble in terms of the first and second derivatives of hypersurface radius $R$ with respect to the proper time defined on it. This, together with an equation of state for the matter on the thin-shell, completely determines the dynamics of the bubble. Stability analysis amounts to consider linearized, radial fluctuations of the bubble around an equilibrium configuration. Let the parameters of such configuration be denoted $\sigma_*$, $p_*$ and $R_*$. Israel conditions yield an effective potential that depends on these parameters and whose concavity permits to decide whether or not the solution is stable, at least under perturbations that preserve the spherical symmetry. It is known that such stable configurations do exist \cite{brady1991stability, visser1995lorentzian}, yielding
\begin{equation}\label{eq:rel_p_sigma_2}
    \frac{p_*}{\sigma_*} = \frac{c^2}{4} \frac{\Delta k}{k_+k_-}\, , \ \ \ \ \Delta k ={k_+ -k_-}\,.
\end{equation}
where $k_+^2 ={ 1 - 1 /R_*}$ and ${k_-^2 = 1 - R_- / R_*}$. From this stability analysis we find that the thin-shell turns out to be stable under radial perturbations if and only if the parameters of the system satisfy the following condition
\begin{align}\label{eq=G}
G(\lambda,\,k_{+},\,k_{-})\,=\,&3(4\,\lambda\,+\,1)\,k_{+}^3\,k_{-}^3\,+\,4\,\lambda\,k_{+}^2\,k_{-}^2\,\nonumber \\
    & -(k_{+}^2\,+\,k_{-}^2\,+\,k_{+}\,k_{-})\,\geq\,0\,,
\end{align}
where $\lambda$ appears in the first order in the expansion of the equation of state $p(\sigma)$; namely, ${dp}/{d\sigma} = c^2 \lambda$. In this way, one interprets ${c \sqrt{\lambda}}$ as the speed of acoustic perturbations on the thin-shell while in equilibrium. The dominant energy condition demands ${\lambda < 1}$. In addition, there are other constraints in the parameter space: Asking for the bubble to be located outside the black hole event horizons and at a distance larger than the Schwarzschild radius of the solution in $\mathcal{M}^+$, i.e. ${R_{\text{ISCO}}>R>R_+=1>R_-}$, implies the parameters ${k_{\pm}}$ to be real. Also, $\sigma_* \geq 0$ is required for the null, the weak and the strong energy conditions to be satisfied. It turns out that this condition is achieved provided $k_-\geq k_+\geq 0$, which in terms of the mass parameters of the Schwarzchild geometries reads ${m_+ \geq m_-}$. All these constraints reduce the region ${G(\lambda,\,k_{+},\,k_{-})} \geq 0$ by a half, cf. \cite{brady1991stability}. In addition, there are constraints on the parameter space coming from stability. Stability under radial perturbations demands the thin-shell to be located at a radius greater than the photon sphere of the inner geometry $\mathcal{M}^-$; namely, ${3R_+>R\,\geq\,1.5\,R_-}$. Still, stable configurations are possible with the thin-shell being either inside or outside the photon sphere of the outer geometry $\mathcal{M}^+$. As we will discuss, the gravitational lensing effect in each of these two scenarios are qualitatively different. As probably expected, the situation is more interesting when $1<R \leq 1.5$, as a second photon sphere forms in this case. This requires, of course, a large pressure $p_* \sim  c^2\sigma_*$, although still compatible with the energy conditions $0\leq p_*<c^2\sigma_*>0$. The minimum values that the equilibrium parameters ${\sigma_*,\,p_*}$ and $\lambda$ can take grow considerably if $R$ is reduced. When the black hole horizon radius, $R_-$, is equal to 1 we have ${k_- = k_+}$ and the shell simply vanishes.

\section{Null geodesics and ray tracing}

In order to obtain the shadow cast by the black hole surrounded by the thin-shell type configuration we have to integrate the geodesic equation on $\mathcal{M}$. This involves some difficulty as the null geodesics in such a stratified geometry may belong to different classes, some of them reentering the bubble and some of them winding around the two possible photon spheres. Besides, when a null geodesic goes through the thin-shell, it experiences a gravitational refraction due to the abrupt change in the space-time curvature, and this ultimately results in a distortion of the shadow we want to characterize. 

Computing the null geodesics in a Schwarzschild patch amounts to consider the conserved quantities $E$, $L$ associated to the Killing vectors $\partial _t$, $\partial _{\phi}$, respectively. We may consider two sets of coordinates $\{t_{\pm},\,r_{\pm},\,\phi_{\pm}\}$, one on each Schwarzschild patch $\mathcal{M}^{\pm }$. In each patch we have constants of motion $L_{\pm},\,E_{\pm}$, respectively. Writing these quantities in terms of a null vector along a given geodesic, and requiring continuity of the metric on $\partial\mathcal{M}^{\pm}$, we find $ L_- = L_+$ and ${E_-}k_+ = {E_+}k_-$. These equations represent the conservation of the angular momentum $L\equiv L_{\pm }$ and the energy red-shift ${E_+}/ {E_-}$ when crossing the thin-shell. Implementing these matching conditions, the ray tracing problem can be solved numerical. We integrate the equation of motion for ${dr_{\pm}/d\phi_{\pm}}$ with fourth order Runge-Kutta. The results can be conveniently expressed in terms of the parameter ${b=c L/E_+}$, which, by studying $dr_{+}/d\phi_{+}$ asymptotically, is identified as the impact parameter. In the standard Schwarzschild geometry with $R_-=R=R_+$, the set of null geodesics can be separated in two classes, depending on whether the value of the impact parameter $b$ is greater or lower than the critical value ${b_c=\sqrt{27}/2}$. 
\begin{figure}
    \centering
    \includegraphics[width=0.42\textwidth]{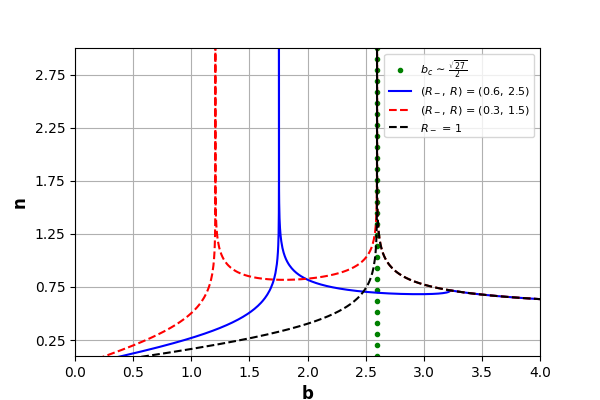}
    \caption{Number of orbits $n$ as a function of the impact parameter $b$. The result for a Schwarzschild black hole is depicted in black. The blue curve corresponds to a configuration with the thin-shell being located outside the photon sphere of the outer geometry. The curved in red corresponds to a configuration with the thin-shell being located at the position of the outer photon sphere. In the latter case, the null geodesics have to photon spheres corresponding to impact parameters ${b_c^{\pm }}$, while in the other two cases the geodesics only have one accumulation value $b^{-}_c$.}
    \label{fig:Vueltas}
\end{figure}
Null geodesics that comes from infinity with $b\geq b_c$ never cross the event horizon, while those with $b< b_c$ unavoidably fall into the black hole. The marginal case ${b\,=\,b_c}$ corresponds to null geodesics winding infinitely many times around the black hole siting in the photon sphere. In the configuration we study here, being a stratified geometry, the classification of null geodesics is notably more involved. While the behaviour of a null geodesic in each Schwarzschild patch will locally be similar, each of the two patches has its own critical impact parameter, namely ${b_c^+ = b_c}$ and ${b_c^-= b_c R_- \sqrt{(R- R_-)/(R-1)}}$.
\begin{figure*}
    \centering
    \includegraphics[width=1.0\textwidth]{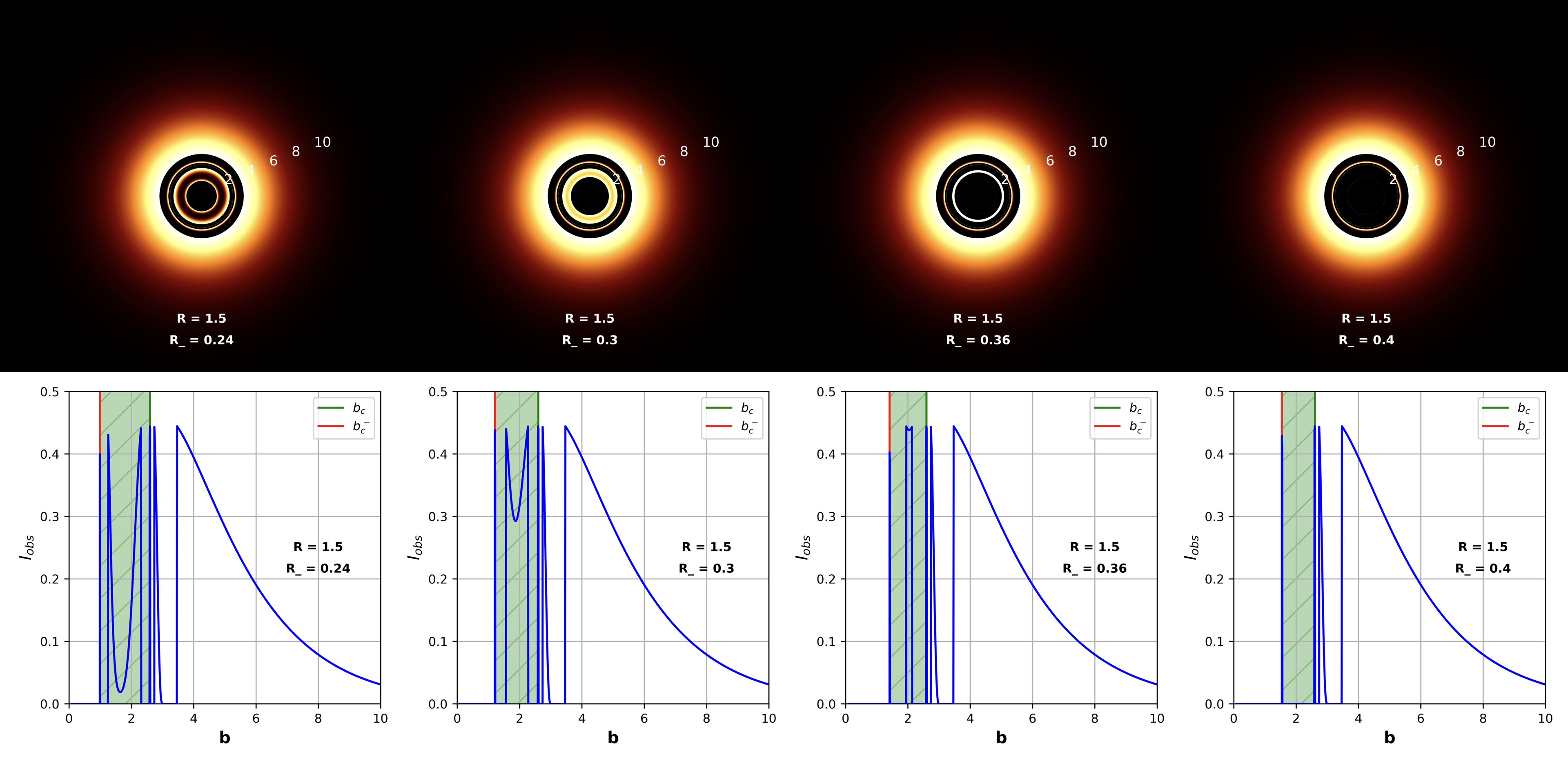}
    \caption{Images produced by configurations with different ratios $R_-/R$, with the thin-shell coinciding with the photon sphere of outer region, i.e. $R=1.5$. Different qualitative behaviors are shown, including single rings, multiple rings, and continuous annulus. Qualitatively similar images occur for ${R \leq 1.5}$. In particular, for ${R = 1.3}$ a continuous ring system associated to the second transfer function is formed for ${R_-\leq 0.18}$; a similar behavior for the third transfer function occurs for $R_-$ between $0.4$ and $0.54$.}
    \label{fig:imagen_R_1.5}
\end{figure*}
Because of this, we have to distinguish among three classes of geodesics: those of class 1 are the null geodesics that only have trajectories in the outer geometry $\mathcal{M}^+$; this happens when $ b>\sqrt{{R^3}/({R-1})}$ if $R>1.5$, or $b>b_c^+$ if $R\leq 1.5$. Null geodesics of class 2 are those that cross the thin-shell and then escape into the outer geometry again; this happens for $b_c^-<b<\sqrt{{R^3}/({R-1})}$ if $R > 1.5$, and for $b_c^-<b<b_c^+ $ if $R\leq 1.5$. The class 3 is defined by the null geodesics that cross the thin-shell and ultimately fall into the black hole, what happens for $b\,<\,b_c^-$.

An important quantity is the total angle deviation of the light ray, ${\Delta \phi}$, which is related to the number of orbits ${n = \Delta\phi / (2 \pi)}$, which is shown in Figure \ref{fig:Vueltas}. $\Delta\phi$ can be expressed as a linear combination of incomplete elliptic integrals of the first kind. Null geodesics belonging to class 1 have the same qualitative behavior as those in the Schwarzschild geometry, and so the integrals can be expanded around the critical impact parameter $b_c$, cf. \cite{luminet1979image, Chandrasekhar:579245}. 
\begin{figure*}
    \centering
\includegraphics[width=0.7\textwidth]{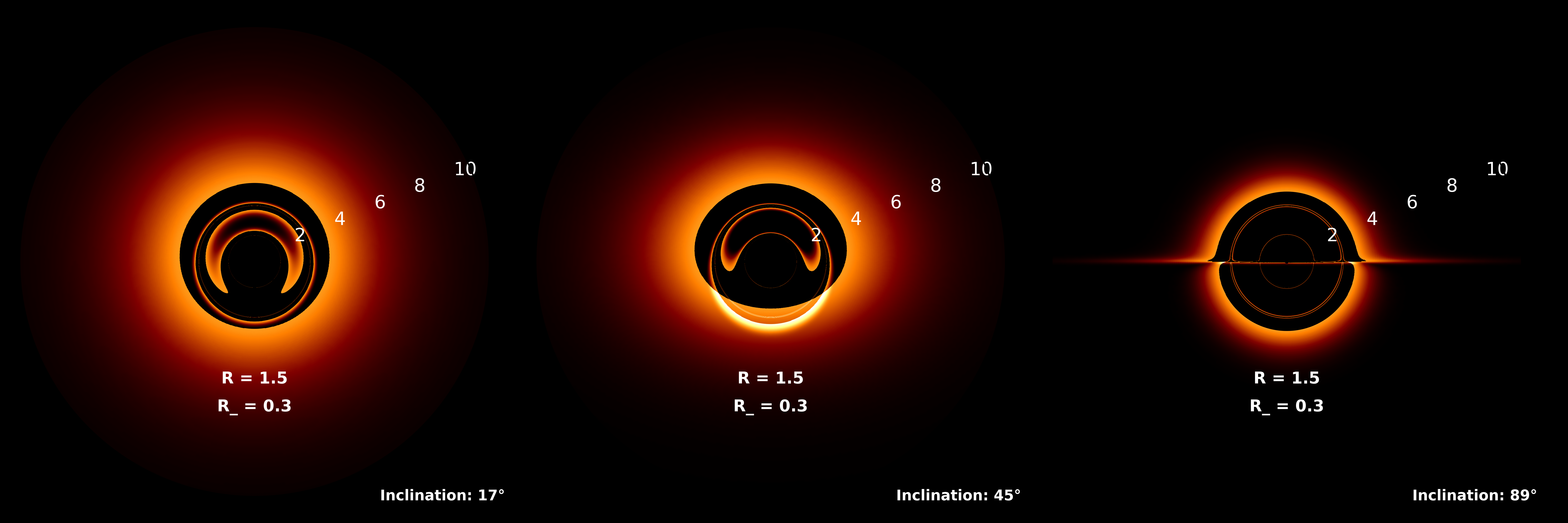}
\includegraphics[width=0.7\textwidth]{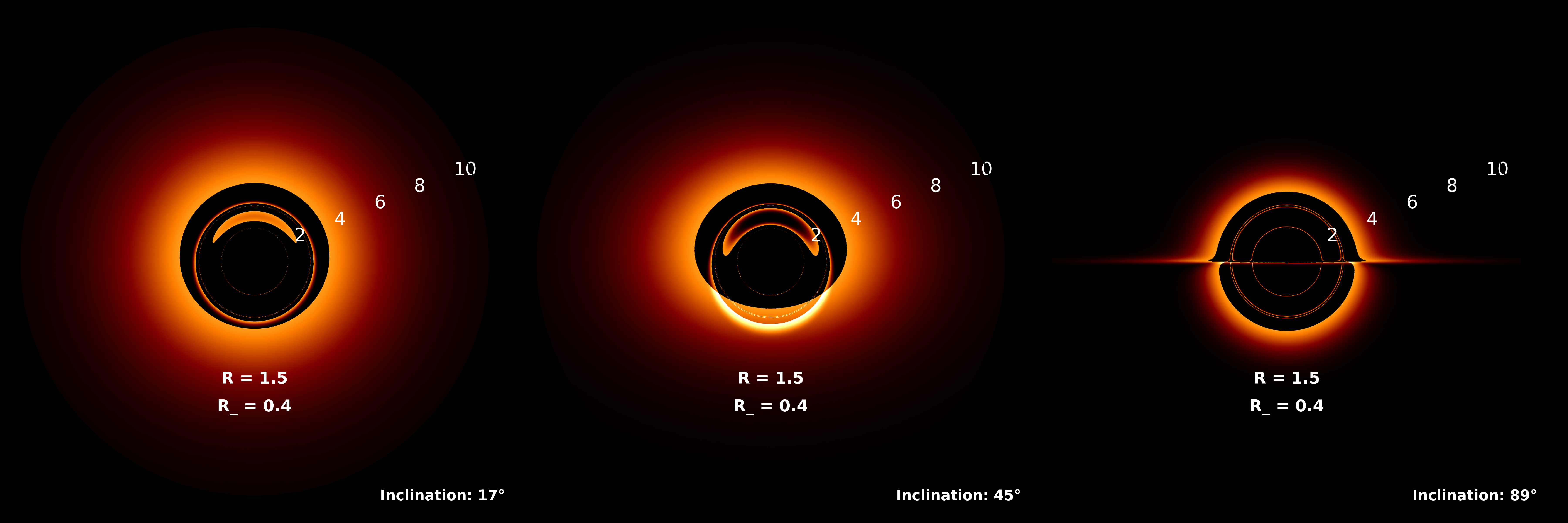}
\includegraphics[width=0.7\textwidth]
{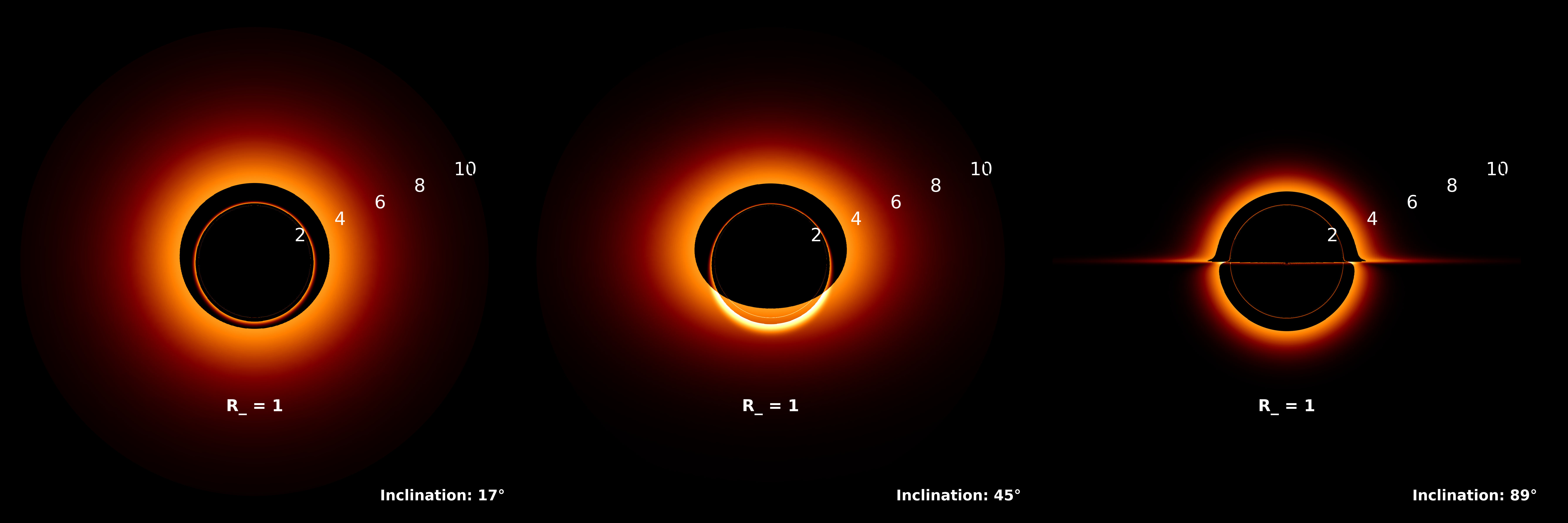}    
\caption{The three images on top (the first line) correspond to ${(R,\,R_-)\,=\,(1.5,\,0.3)}$ with inclination angles 90º$-\alpha = \{ $17º, 45º, 89º$\}$, respectively. Images in the second line correspond to ${(R,\,R_-)\,=\,(1.5,\,0.4)}$, with the same sequence of inclination angle. Images in the third line show the case of a Schwarzschild black hole with no bubble ($R_-=1$) and is included for comparison; the same angles are presented in this case.}
\label{fig:inclinado_0.3.png}
\end{figure*}
Something similar can be done for the geodesics belonging to the classes 2 and 3; expansions of $\Delta\phi$ can be obtained around each critical values $b_c=b_c^+$ and $b_c^-$. Defining the variation of the impact parameter as ${b=b^{\pm }_c+ \delta b^{\pm}}$, it can be seen from the expansions of $\Delta\phi$ that the dependence is exponential, namely $|{\delta b^{\pm}|\sim e^{-\Delta\phi}}$. This turns out to be useful to characterize the annular and sub-annular structure in the image produced by the entire configuration and sourced by the surrounding matter --the disk--. In the next section, we will model the accretion disk and obtain the exact shape of the shadow.

\section{Accretion disk}\label{ch_disco_acreción}

We consider an optically and geometrically thin accretion disk concentric with the system. For clarity, we will first consider the polar view ($\alpha=$ 90º in Figure \ref{fig:esquema.png}) and later study the most general configuration with arbitrary inclination angle. Doing so will enable us to distinguish among different effects. We model the disk in a standard way: we assume that its intensity $I_{(\nu )}$ for a given frequency $\nu$ only depends on the radial coordinate, and that its emission is isotropic in the frame that is at rest with matter. Then, we assume that the disk emits as a black body, which enables us to integrate over the frequencies and write the observed intensity $I^{\text{obs}}$ in terms of the emitted intensity $I^{\text{em}}$. This yields 
\begin{equation}\label{arribeno}
{I^{\text{obs}} = \left(\frac{\nu_{\text{obs}}}{\nu_{\text{em}}}\right)^4\, I^{\text{em}}}\,  .
\end{equation}
The factor $g{(r_{\text{em}})}\equiv {\nu_{\text{obs}}}/{\nu_{\text{em}}}$ measures the redshift relative to an observer ($\text{obs}$) that will ultimately be considered at infinity. In deriving (\ref{arribeno}) we used that, along a ray, $\delta{I_{({\nu})} \sim \nu^3\delta \nu}$, cf. \cite{frank2002accretion}. As the accretion disk is optically thin, we can neglect absorption. Under this hypothesis, the variation of $I_{(\nu )}$ in the medium will only depend on the emission coefficient $j_{\nu}=\hat{n} \cdot \nabla I_{(\nu )}$, which is the density of energy emitted per unit of time, per unit of solid angle, in a given direction $\hat n$, and for a given frequency $\nu$.

When computing the (inverse) ray tracing, one has to consider that every time that a light ray intersects the accretion disk picks up brightness from the disk emission. So the observed intensity is given by 
\begin{equation}
I^{\text{obs}}_{(b)} = \sum_{m\geq 1} g^4_{({r_m(b)})}\, I^{\text{em}}_{({r_m(b)})}\,,
\end{equation}
where ${r_m}(b)$ is the so-called transfer function, and corresponds to the radial coordinate of the $m^{\text{th}}$ encounter between the disk and the null geodesic that at infinity has impact parameter $b$, \cite{PhysRevD.100.024018}; namely, $r_m(b) = r(\phi_m)$ with $\phi_m = (m - \frac 12){\pi}$. As the luminosity profile of the disk, we have proven different functions. A detailed analysis of these functions has been done in \cite{PhysRevD.100.024018} for other scenarios; see also \cite{zeng2020influence}. It turns out that, for a scenario like ours, different emission profiles lead to similar effects. Therefore, it is enough to consider the example of a profile that decays exponentially with the distance from the ISCO.

\section{Sub-annular images}

Producing the shadow image amounts to consider the contribution of multiple images and the respective magnification. This implies to take into account the relative contribution of the different classes of geodesics that experience $m$ encounters with the disk, and to compare the contribution of the leading cases $m=1, 2$ relative to that of the subdominant ones $m\geq 3$. This analysis is sensitive to the choice of the parameters $R,\, R_-$ of the configuration; still, it can be done systematically: We denote by ${b^{\pm}_m}$ the edges of the image of $m^{\text{th}}$ order. That is, such image will correspond to values of the impact parameter within the interval $(b_m^-,\,b_m^ +)$. The null geodesics that cross the disk $m$ times have angular deviation $\Delta \phi$ between $(m-\frac 12)\pi$ and $(m+\frac 12)\pi$. As the variation of the impact parameter exhibits an exponential dependence with $\Delta\phi$, the width of each ring associated with the image of order $m^{\text{th}}$ will decay exponentially, yielding ${\Delta b_m \equiv b_m^+-b_m^-\sim e^{-\pi} \Delta b_{m - 1}}$. This implies that the images are exponentially suppressed, while superposed on the ring pattern they contribute to form. A systematic inspection leads to the conclusion that is natural to set a cutoff at $m=4$, cf. \cite{PhysRevD.100.024018}. 

The analysis of the different geodesics and multiple images permits to analyze the shadow cast by different dynamical configurations. In the region of the parameter space that corresponds to the thin-shell being located at a radius $R>1.5$, the shadow is expected to be qualitatively similar to that of a Schwarzschild black hole, the reason being that in that case there is only one photon sphere, resulting in a single photon ring whose radius will depend on $R_-$. An interesting situation is when the thin-shell lies inside the photon sphere of the outer region $\mathcal{M}^+$. Naively, this would lead to the presence of two sets of rings in the image, one for each photon sphere. However, while in same cases this is what actually happens, in general one has to perform a careful analysis of the average width and intensity of images of different orders. Whether or not a ring pattern is observable in the final image depends on the interplay between the transfer functions associated to the different orders ${m = 1, \,2,\, 3,...}$ as well as on the radial cut of accretion disk. A case by case analysis of this is possible, but summarizing it here would not be more illuminating than taking a look at Figure \ref{fig:imagen_R_1.5} and the explanation in its epigraph. There, we observe that a multiple rings pattern is possible for certain range of parameters. It is also possible to have a continuum band in the ring structure. The most interesting phenomenon is the existence of sub-annular structure inside the photon ring. The ring pattern becomes even motlier once we consider a generic inclination angle $\alpha$ of the accretion disk. For arbitrary inclination, the transfer functions have to be defined as $r_m(b) = r(\phi_m)$ where $\phi_m = (m -\frac 12)\pi\pm\beta(\varphi)$, and $\cos^2(\beta)\,=\,{\left(1\,+\,\sin^{2}(\varphi)\,\tan^2(\alpha)\right)^{-1}}$, with $\varphi $ being the angle that parameterizes the disk, while $\phi $ is the polar angle from the observer viewpoint. Different inclination angles are shown in Figure \ref{fig:inclinado_0.3.png}, where one can clearly observe the formation of sub-annular structures due to the gravitational refraction effect.

\section{Conclusions}

In this work we have studied a new type of phenomenon that can occur in black hole imaging. This is the possibility that a certain type of concentration of non-emitting matter around a black hole produces gravitational refraction that gives rise to sub-annular images in the shadow of black holes; that is, images within the so-called photon ring. The interesting thing about this phenomenon is the possibility of having sub-annular images in a scenario that does not require the inclusion of exotic matter or strange space-time contortions, but rather a concentration of --dark-- matter that, forming a stable structure, satisfies all the energy conditions. Something we have not discussed here is the nature of this non-emitting matter. To form the configuration that we have considered, this matter must self-interact at the level of presenting an equation of state with pressures that, while obeying the energy conditions, is relativistic. We will not enter here into speculations about what the nature of the supposed relativistic, self-interacting dark matter that forms the bubble may be. However, it is worth saying that, as long as one avoids considering that it is a main component of dark matter, it is totally sensible to consider the presence of such type of matter and investigate the possible phenomena it might produce. More general density profiles of the matter in the bubble are expected to produce optical effects qualitatively similar to those described here.


\end{document}